\documentclass[prd,aps]{revtex4}
\usepackage{amssymb}
\newcommand{\Real}{\mathbb{R}}

\begin{document}
%%%%%%%%%%%%%%%%%%%%%%%%%%%%%%%

\title{On the well posedness of the Baumgarte-Shapiro-Shibata-Nakamura
formulation of Einstein's field equations}
\author{Horst Beyer}
\affiliation{Department of Mathematics, Louisiana State University,
Lockett Hall, Baton Rouge, Louisiana 70803-491, USA; Center for
Computation \& Technology, Frey Building, Baton Rouge, LA 70803, USA
and MPI for Gravitational Physics, Am Muehlenberg 1, 14476 Golm,
Germany.}
\author{Olivier Sarbach}
\affiliation{Department of
Mathematics, Louisiana State University, Lockett Hall, Baton Rouge,
Louisiana 70803-491, USA, and
Department of Physics and Astronomy, Louisiana State University, 202
Nicholson Hall, Baton Rouge, Louisiana 70803-4001.}

\begin{abstract}
We give a well posed initial value formulation of the
Baumgarte-Shapiro-Shibata-Nakamura form of Einstein's equations with
gauge conditions given by a Bona-Mass{\'o} like slicing condition for
the lapse and a frozen shift. This is achieved by introducing extra
variables and recasting the evolution equations into a first order
symmetric hyperbolic system. We also consider the presence of
artificial boundaries and derive a set of boundary conditions that
guarantee that the resulting initial-boundary value problem is well
posed, though not necessarily compatible with the constraints. In the
case of dynamical gauge conditions for the lapse and shift we obtain a
class of evolution equations which are strongly hyperbolic and so
yield well posed initial value formulations.
\end{abstract}

\maketitle

\section{Motivation}

Most numerical evolutions of Einstein's field equations try to
approximate solutions on a generically infinite (non-compact) 3-space
by computations on a truncated finite (compact) domain. For this,
artificial boundaries and corresponding boundary conditions have to be
introduced. Mathematically, this immediately poses the question of
well-posedness of the initial-boundary-value problem (IBVP) for the
evolution equations and compatibility with the constraints. In
addition, governed by causality, the solution on the finite domain is
very likely to differ from the solution on the infinite domain, after
disturbances from the boundaries enter the computational domain. This
makes the choice of the boundary condition crucial for the physical
interpretation of the results, especially if one thinks of integrated
quantities like masses, charges and momenta. Physically, it could be
even argued that ultimately only estimates of the deviation of the
numerical solution from the actual solution are significant and that
the mathematical concepts of well-posedness of the IBVP including
compatibility with the constraints and avoidance of reflections from
artificial boundaries are only steps towards achieving this goal.
\newline
\linebreak Removing the influence of the boundaries could be achieved
by enlarging the computational domain to a size such that, according
to causality, disturbances from boundaries cannot have reached the
domain of physical interest. Note that this requires knowledge of the
causality structure of the spacetime, which presupposes estimates on
the solution to be found. In addition, increasing the size of the
computational domain goes at the cost of resolution, because of finite
computational resources, which is particularly restricting in
$3-$dimensional problems although this problem can be alleviated by
using adaptive or fixed mesh refinement techniques \cite{AMR}. So,
both restrictions in computational resources and the demand for higher
resolution, lead us to attempt to minimize the influence of artificial
boundaries on the numerical solution. This is tried by so called
`outgoing boundary conditions' meant to make those boundaries to
appear as `transparent' as possible. For instance, one such approach
is given by Endquist and Majda \cite{EM} using a hierarchy of
conditions which gradually decrease reflections at the
boundary. Alternative approaches are the methods of characteristic
\cite{C-Match} or perturbative matching \cite{P-Match}. See
\cite{Helmut, Frauendiener} for an approach trying to avoid the
introduction of artificial boundaries altogether by a suitable
compactification of spacetime.

\section{Introduction}

In this article we analyze the IBVP of the
Baumgarte-Shapiro-Shibata-Nakamura (BSSN) \cite{SN, BS} formulation of
Einstein's vacuum equations, which is currently used by several groups
in numerical relativity with applications to the binary black hole and
binary neutron star problem, see \cite{BSSN} for a review. Since the
BSSN equations are first order in time, but mixed first/second order
in space, their type (elliptic, parabolic, hyperbolic or mixed) is a
priori not clear. Here, we analyze the well posedness of their
(nonlinear) Cauchy problem with and without boundaries. Well posedness
means that the Cauchy problem has a unique solution local in time and
that the solution depends continuously on the initial data. The last
property is important in view of obtaining convergent discretizations
since in general numerical simulations introduce small errors in the
initial data. If violated, this can lead to errors at a later time
which grow exponentially with increasing resolution (see
\cite{GKO-Book, CPST-Convergence} for examples of this phenomenon).
\newline
\linebreak We find that the BSSN system with a large family of gauge
conditions for the lapse, including Bona-Mass{\'o} like slicing
conditions, and an a priori specified shift yields a well posed initial
value problem. This is achieved by introducing extra fields that make
it possible to recast the system into a first order quasi-linear
symmetric hyperbolic form for which standard well posedness results
are known \cite{Kato}. The introduction of extra fields brings
additional constraints, and the original BSSN system and the first
order symmetric hyperbolic system derived in this article are only
equivalent if these constraints are satisfied. However, we show that
the associated constraint variables obey a closed evolution system
that is {\em independent} of the other constraints. This means that
the additional constraints are satisfied everywhere at later times if
satisfied initially, even if the other constraints are violated. This
implies that the (original) BSSN system is well posed; in particular,
unique solutions local in time exist, and depend continuously on the
initial data.
\newline
\linebreak Our first order symmetric hyperbolic reduction also
facilitates the analysis of characteristic modes which is particularly
useful when constructing boundary conditions. Here we construct
maximally dissipative boundary conditions that guarantee the well
posedness of the resulting IBVP \cite{MaxDiss}. Although in general
these boundary conditions are not compatible with the constraints,
they are consistent with the evolution equations and constitute a
first step towards improving numerical evolutions of the BSSN
system. In particular, the present analysis offers the possibility to
construct constraint-preserving boundary conditions \cite{CPBC} in the
linearized case, following the lines of \cite{CPSTR, CS, GMG2}.
\newline
\linebreak The techniques used in this article are the same used in
\cite{LSU-BSSN} where well posedness of the BSSN system with an
explicitly given shift and an algebraic gauge condition is found by
considering an auxiliary first order system. A different technique
which makes use of pseudo-differential calculus has recently been
applied in order to show well posedness for a closely related
formulation \cite{NOR}. More recently, in \cite{GMG1, GMG2} a
definition of symmetric hyperbolicity based on energy estimates for
second order systems was presented which was verified for the BSSN
system and the formulation in \cite{NOR} for the case of an algebraic
lapse and an explicitly given shift. Nevertheless the connection of
their definition and existence of solutions is open.
\newline
\linebreak The remainder of this work is organized as follows. In
section \ref{Sect:BSSNEq} we review the BSSN equations, specify the
gauge conditions we are considering and discuss the evolution system
for the constraint variables. In section \ref{Sect:FOSH} we introduce
extra fields and derive a first order symmetric hyperbolic system that
reflects the dynamics of the original BSSN system. The characteristic
fields with nontrivial speeds are constructed in section \ref{Sect:BC}
and are used to write down maximally dissipative boundary
conditions. In section \ref{Sect:LiveShift} we find using
pseudo-differential calculus that the BSSN system with a
``$K$-driver'' and a ``Gamma-freezing'' condition as defined in
\cite{AEI} but with a different time coordinate is strongly hyperbolic
according to the definition in \cite{KO, NOR} and so yields a well
posed initial value formulation. Conclusions are drawn in section
\ref{Sect:Conclusion}.

\section{The BSSN equations}
\label{Sect:BSSNEq}

Since our results depend crucially on the principal part of the
equations, we write down the BSSN system explicitly in this section.
The system of equations is the one that has been used in
Ref. \cite{AEI} for numerical simulations, but it might differ from
the one used by other groups. Decomposing the three metric and the
extrinsic curvature according to
\begin{eqnarray}
\gamma_{ij} &=& e^{4\phi}\tilde{\gamma}_{ij}\; ,\\
K_{ij} &=& e^{4\phi}\left( \tilde{A}_{ij} + \frac{1}{3}\tilde{\gamma}_{ij} K \right),
\end{eqnarray}
where $\tilde{\gamma}_{ij}$ has unit determinant and $K = \gamma^{ij}
K_{ij}$ is the mean curvature, the evolution equations are obtained from
\begin{eqnarray}
\hat{\partial}_0 \phi &=& -\frac{\alpha}{6}\, K + \frac{1}{6}\partial_k\beta^k, 
\label{Eq:BSSN1}\\
\hat{\partial}_0 \tilde{\gamma}_{ij} &=& -2\alpha\tilde{A}_{ij} 
 + 2\tilde{\gamma}_{k(i}\partial_{j)}\beta^k - \frac{2}{3}\tilde{\gamma}_{ij}\partial_k\beta^k ,
\label{Eq:BSSN2}\\
\hat{\partial}_0 K &=& -e^{-4\phi}\left[ \tilde{D}^i\tilde{D}_i \alpha - 2\partial_i\phi \cdot\tilde{D}^i\alpha \right]
 + \alpha\left( \tilde{A}^{ij}\tilde{A}_{ij} + \frac{1}{3} K^2 \right)
 - \alpha S,
\label{Eq:BSSN3}\\
\hat{\partial}_0 \tilde{A}_{ij} &=& 
 e^{-4\phi}\left[ \alpha\tilde{R}_{ij} + \alpha R^\phi_{ij} - \tilde{D}_i\tilde{D}_j\alpha 
  - 4\partial_{(i}\phi\cdot\tilde{D}_{j)}\alpha\right]^{TF}
\nonumber\\
 &+& \alpha K\tilde{A}_{ij} - 2\alpha\tilde{A}_{ik}\tilde{A}^k_{\; j}
  + 2\tilde{A}_{k(i}\partial_{j)}\beta^k - \frac{2}{3}\tilde{A}_{ij}\partial_k\beta^k
  - \alpha e^{-4\phi} \hat{S}_{ij} ,
\label{Eq:BSSN4}\\
\hat{\partial}_0\tilde{\Gamma}^i &=& \tilde{\gamma}^{kl}\partial_k\partial_l\beta^i
 + \frac{1}{3} \tilde{\gamma}^{ij}\partial_j\partial_k\beta^k 
 + \partial_k\tilde{\gamma}^{kj} \cdot \partial_j\beta^i
 - \frac{2}{3}\partial_k\tilde{\gamma}^{ki} \cdot \partial_j\beta^j\nonumber\\
 && - 2\tilde{A}^{ij}\partial_j\alpha + 2\alpha\left[ (m-1)\partial_k\tilde{A}^{ki} - \frac{2m}{3}\tilde{D}^i K
    + m(\tilde{\Gamma}^i_{\; kl}\tilde{A}^{kl} + 6\tilde{A}^{ij}\partial_j\phi) \right]
    - S^i,
\label{Eq:BSSN5}
\end{eqnarray}
where we have introduced the operator $\hat{\partial}_0 = \partial_t -
\beta^j\partial_j$. Here, all quantities with a tilde refer to the
conformal three metric $\tilde{\gamma}_{ij}$, and the latter is used
in order to raise and lower their indices. In particular,
$\tilde{D}_i$ and $\tilde{\Gamma}^k_{\; ij}$ refer to the covariant
derivative and the Christoffel symbols, respectively, with respect to
$\tilde{\gamma}_{ij}$. The expression $[ ... ]^{TF}$ denotes the
trace-less part (with respect to the metric $\tilde{\gamma}_{ij}$) of
the expression inside the parentheses, and
\begin{eqnarray}
\tilde{R}_{ij} &=& -\frac{1}{2} \tilde{\gamma}^{kl}\partial_k\partial_l\tilde{\gamma}_{ij} 
  + \tilde{\gamma}_{k(i}\partial_{j)}\tilde{\Gamma}^k
  - \tilde{\Gamma}_{(ij)k}\partial_j\tilde{\gamma}^{jk} 
  + \tilde{\gamma}^{ls}\left( 2\tilde{\Gamma}^k_{\; l(i}\tilde{\Gamma}_{j)ks} + \tilde{\Gamma}^k_{\; is}\tilde{\Gamma}_{klj} \right),
\nonumber\\
R^\phi_{ij} &=& -2\tilde{D}_i\tilde{D}_j\phi - 2\tilde{\gamma}_{ij} \tilde{D}^k\tilde{D}_k\phi
  + 4\tilde{D}_i\phi\, \tilde{D}_j\phi - 4\tilde{\gamma}_{ij}\tilde{D}^k\phi\, \tilde{D}_k\phi.
\nonumber
\end{eqnarray}
The parameter $m$, which was introduced in \cite{Alc-Stab},
controls how the momentum constraint is added to the evolution
equations for the variable $\tilde{\Gamma}^i$. The system in
Ref. \cite{AEI} corresponds to the choice $m=1$. However, in order to
obtain a first order symmetric hyperbolic reduction, we will see later
that we need $m$ to be a specific function of the lapse and the mean
curvature. The source terms $S$, $\hat{S}_{ij}$ and $S^i$ are defined
in terms of the four Ricci tensor, $R^{(4)}_{ij}$, and the constraint
variables
\begin{eqnarray}
H &\equiv& \frac{1}{2}\left( \gamma^{ij} R^{(3)}_{ij} + K^2 - K^{ij} K_{ij} \right),
\label{Eq:BSSNCons1}\\
M_i &\equiv& \tilde{D}^j \tilde{A}_{ij} - \frac{2}{3} \tilde{D}_i K + 6\tilde{A}_{ij} \tilde{D}^j\phi,
\\
C^i_\Gamma &\equiv& \tilde{\Gamma}^i + \partial_j\tilde{\gamma}^{ij}\; ,
\label{Eq:BSSNCons3}
\end{eqnarray}
as
\begin{eqnarray}
S &=& \gamma^{ij} R^{(4)}_{ij} - 2 H,\\
\hat{S}_{ij} &=& \left[ R^{(4)}_{ij} + \tilde{\gamma}_{k(i}\partial_{j)} C_\Gamma^k \right]^{TF},\\
S^i &=& 2\alpha\, m\,\tilde{\gamma}^{ij} M_j - \hat{\partial}_0 C_\Gamma^i\; .
\end{eqnarray}
The vacuum equations consist of the evolution equations (\ref{Eq:BSSN1}-\ref{Eq:BSSN5})
with $S=0$, $\hat{S}_{ij}=0$, $S^i = 0$ and the constraints $H=0$, $M_i = 0$ and
$C_\Gamma^i = 0$.

Using the Bianchi identities, $2\nabla^\mu R^{(4)}_{\mu\nu} -
\nabla_\nu R^{(4)} = 0$ and imposing the evolution equations, it can
be shown that the constraint variables obey the following propagation
system:
\begin{eqnarray}
\hat{\partial}_0 H &=& -\frac{1}{\alpha}\, D^j(\alpha^2 M_j) 
 - \alpha e^{-4\phi}\tilde{A}^{ij}\tilde{\gamma}_{ki}\partial_j C^k_\Gamma + \frac{2\alpha}{3}\, K H,
\label{Eq:H}\\
\hat{\partial}_0 M_j &=& \frac{\alpha^3}{3} D_j( \alpha^{-2} H ) + \alpha K M_j 
 + D^i\left( \alpha\left[ \tilde{\gamma}_{k(i}\partial_{j)}C^k_\Gamma \right]^{TF} \right),
\label{Eq:Mj}\\
\hat{\partial}_0 C^k_\Gamma &=& 2\alpha\, m\, \tilde{\gamma}^{kl} M_l\; .
\label{Eq:Ck}
\end{eqnarray}
By introducing the further constraint variable $Z_j^{\;\; k} =
\partial_j C^k_\Gamma$ which satisfies $\partial_{[i} Z_{j]}^{\;\; k} =
0$ one can reduce Eqs. (\ref{Eq:H},\ref{Eq:Mj},\ref{Eq:Ck}) to a first
order symmetric hyperbolic system provided that $m > 1/4$. In the
absence of boundaries, this implies that the constraints are
preserved, i.e. trivial initial data for the constraints variables
lead to zero constraint variables at later times as well. If time-like
boundaries are present, the constraints are only preserved if suitable
boundary conditions are specified. Such constraint-preserving boundary
conditions are discussed in \cite{CPBC, CPSTR, CS, GMG2}; but for the
(nonlinear) BSSN system it is not yet understood if they lead to a
well posed IBVP.

In order to evolve the system (\ref{Eq:BSSN1}-\ref{Eq:BSSN5}) we have
to specify conditions on the lapse $\alpha$ and the shift $\beta^i$.
The simplest possibility is to set $\alpha=1$ (or any other fixed
function) and $\beta^i=0$. However, this leads to a formulation that
is not strongly hyperbolic (this will follow from the results in
Sect. \ref{Sect:LiveShift} if we set the function $f$ defined below in
Eq. (\ref{Eq:LapseAEI}) to zero). This can be avoided by
``densitizing'' the lapse. More generally, we can require
\cite{ST-Live} that the lapse
\begin{equation}
\alpha = \alpha(\phi,x^\mu),
\end{equation}
is a smooth strictly positive function of the conformal factor (or the
determinant of the three metric) and spacetime coordinates with the
restriction that $\sigma = (12\alpha)^{-1}\partial\alpha/\partial\phi$
is strictly positive. Taking a time derivative of this, assuming that
$\partial\alpha/\partial t = 0$ and using Eq. (\ref{Eq:BSSN1}) we
obtain
\begin{equation}
\frac{d}{dt}\, \alpha = -2\alpha^2\sigma\left( K - \frac{1}{\alpha} D_k\beta^k \right),
\end{equation}
which is the modification of the Bona-Mass{\'o} condition
\cite{BM-Live} proposed in \cite{Ollin1, Ollin2}. The advantage of
this gauge is that it is compatible with a time-independent lapse in a
time slicing that is adapted to stationarity if $\partial_t$ is a
Killing field. It follows from the calculations of
Ref. \cite{LSU-BSSN} that in this case the BSSN system is strongly
hyperbolic if one chooses $m > 1/4$ and symmetric hyperbolic if the
parameter $m$ is adjusted such that $4m = 6\sigma + 1$ with $\sigma >
1/2$\footnote{In \cite{GMG2} it is shown that the system is still
symmetric hyperbolic according to their definition if one extends the
parameter space to $4m = 6\sigma + 1$ with $\sigma\geq 1/2$.}. We
mention here that the special case $\alpha = e^{6\phi} Q(x^\mu)$,
where $Q(x^\mu)$ is an a priori specified function, has been observed
to lead to more stable numerical evolutions of single black hole with
the BSSN system \cite{Pablo}.

Here, we are interested in live gauge conditions which allow lapse and
shift to react on changes of the fields. Such conditions can be
useful, for instance, to evade singularities. In this article, we
consider two cases of gauge conditions:
\begin{enumerate}
\item[(a)]
The following evolution equation for the lapse
\begin{equation}
\hat{\partial}_0 \alpha = -\alpha F(\alpha,K,x^\mu),
\label{Eq:Lapse}
\end{equation}
where $F$ is a smooth function of $\alpha$, $K$ and $x^\mu$
with the restriction that
\begin{equation}
\sigma \equiv \frac{1}{2\alpha}\frac{\partial F}{\partial K} > 0.
\end{equation}
This condition generalizes the Bona-Mass{\'o} gauges. The shift is
frozen, that is, assumed to be an a priori specified function of
spacetime. Symmetric hyperbolic formulations of the vacuum field
equations with these gauge conditions were obtained in \cite{ST-Live}.
\item[(b)]
The gauge conditions of Ref. \cite{AEI} which, for the
lapse, require the ``hyperbolic $K$-driver'' condition
\begin{equation}
\hat{\partial}_0 \alpha = -\alpha^2 f(\alpha,\phi,x^\mu) (K - K_0(x^\mu)),
\label{Eq:LapseAEI}
\end{equation}
where the function $f(\alpha,\phi,x^\mu)$ is smooth and strictly
positive, and $K_0(x^\mu)$ is an arbitrary smooth function. For the
shift, the ``hyperbolic Gamma driver'' \cite{AEI} type condition
\begin{eqnarray}
\hat{\partial}_0 \beta^i &=& \alpha^2 G(\alpha,\phi,x^\mu) B^i,
\label{Eq:BetaAEI1}\\
\hat{\partial}_0 B^i &=& e^{-4\phi} H(\alpha,\phi,x^\mu)\hat{\partial}_0\tilde{\Gamma}^i - \eta(B^i,\alpha,x^\mu)
\label{Eq:BetaAEI2}
\end{eqnarray}
is imposed, where $G(\alpha,\phi,x^\mu)$ and $H(\alpha,\phi,x^\mu)$
are smooth, strictly positive function, and $\eta(B^i,\alpha,x^\mu)$
is a smooth function. Notice that Eq. (\ref{Eq:LapseAEI}) is a special
case of Eq. (\ref{Eq:Lapse}). Note also that the conditions
(\ref{Eq:LapseAEI},\ref{Eq:BetaAEI1},\ref{Eq:BetaAEI2}) differ from
the ones considered in \cite{AEI} by the replacement $\partial_t
\mapsto \hat{\partial}_0$ which simplifies the analysis in the present
article.
\end{enumerate}

In the next section, we show that the gauge conditions (a) lead to a
well posed initial value problem provided that the parameter $m$ is
chosen such that $4m = 6\sigma + 1$. In the presence of boundaries, we
derive boundary conditions in section (\ref{Sect:BC}) that make sure
that in this case the resulting IBVP is well posed. In section
(\ref{Sect:LiveShift}) we show that the initial value problem with the
gauge conditions (b) is well posed provided that some specified
conditions on $m$ and the functions $f$, $G$ and $H$ are
satisfied. Symmetric hyperbolic first order formulations of Einstein's
equations that incorporate gauge conditions that are similar to (b)
have been worked out in \cite{LS-Live}.

\section{First order symmetric hyperbolic form (frozen shift)}
\label{Sect:FOSH}

In this section we recast the BSSN equations with the gauge conditions
(a) into a first order symmetric hyperbolic system. In order to do so
we introduce the extra variables
\begin{equation}
d_k = 12\partial_k\phi, \;\;\;
\tilde{d}_{kij} = \partial_k\tilde{\gamma}_{ij}\, ,\;\;\;
A_k = \frac{\partial_k\alpha}{\alpha}\; ,
\label{Eq:ExtraVars}
\end{equation}
and rewrite Eqs. (\ref{Eq:BSSN3},\ref{Eq:BSSN4},\ref{Eq:BSSN5}) as
\begin{eqnarray}
\hat{\partial}_0 K &=& -\alpha e^{-4\phi}\tilde{\gamma}^{ij}\partial_i A_j + l.o.,
\label{Eq:BSSN3fo}\\
\hat{\partial}_0 \tilde{A}_{ij} &=& \alpha e^{-4\phi}\left[ 
 -\frac{1}{2}\tilde{\gamma}^{kl}\partial_k\tilde{d}_{lij} 
 + \zeta\tilde{\gamma}^{kl} C^{\tilde{d}}_{k(ij)l} 
 + \tilde{\gamma}_{k(i}\partial_{j)}\tilde{\Gamma}^k
 - \frac{1}{6} \partial_{(i} d_{j)} - \partial_{(i} A_{j)} \right]^{TF}
 + l.o.,
\label{Eq:BSSN4fo}\\
\hat{\partial}_0\tilde{\Gamma}^i &=& 
  2\alpha\left[ (m-1)\partial_k\tilde{A}^{ki} - \frac{2m}{3}\tilde{D}^i K \right]
  + l.o.,
\label{Eq:BSSN5fo}
\end{eqnarray}
where $l.o.$ refers to lower order terms that depend on $\phi$,
$\tilde{\gamma}_{ij}$, $K$, $\tilde{A}_{ij}$, $\tilde{\Gamma}^i$,
$\alpha$, $d_k$, $\tilde{d}_{kij}$, $A_k$ but not their
derivatives. Here, we have added the constraint variables
$C^{\tilde{d}}_{lkij} = \partial_{[l}\tilde{d}_{k]ij}$ with an
arbitrary parameter $\zeta$ in the equation for $\tilde{A}_{ij}$. As
we will see shortly, the addition of these constraints will allow us
to obtain a larger family of symmetric hyperbolic formulations.
Evolution equations for the extra variables are obtained by applying
the operator $\hat{\partial}_0$ on the equations (\ref{Eq:ExtraVars}),
using the commutation relation $[\hat{\partial}_0,\partial_k] =
\partial_k\beta^l\cdot\partial_l$ and using the evolution equations
(\ref{Eq:BSSN1},\ref{Eq:BSSN2},\ref{Eq:Lapse}) for $\phi$,
$\tilde{\gamma}_{ij}$ and $\alpha$. The result is
\begin{eqnarray}
\hat{\partial}_0 d_k &=& -2\alpha(\partial_k + A_k)K 
 + d_l\partial_k\beta^l + 2\partial_k\partial_l\beta^l, 
\label{Eq:BSSN8}\\
\hat{\partial}_0 \tilde{d}_{kij} &=& -2\alpha(\partial_k + A_k)\tilde{A}_{ij} 
   + \tilde{d}_{lij}\partial_k\beta^l
  + 2\tilde{d}_{kl(i}\partial_{j)}\beta^l -\frac{2}{3}\tilde{d}_{kij}\partial_l\beta^l
  + 2\tilde{\gamma}_{l(i} \partial_{j)}\partial_k\beta^l 
  - \frac{2}{3} \tilde{\gamma}_{ij} \partial_k\partial_l\beta^l,
\label{Eq:BSSN9}\\
\hat{\partial}_0 A_k &=& -2\sigma \alpha\partial_k K - \alpha\frac{\partial F}{\partial\alpha} A_k 
  - \frac{\partial F}{\partial x^k} + A_l\partial_k\beta^l \, .
\label{Eq:BSSN10}
\end{eqnarray}

We have rewritten the BSSN equations (with a fixed prescribed shift
but a live condition for the lapse) as a first order quasi-linear
evolution system for the variables $u =
(\phi,\tilde{\gamma}_{ij},\alpha,K,\tilde{A}_{ij},\tilde{\Gamma}^k,d_k,\tilde{d}_{kij},A_k)^T$
which is given by the equations
(\ref{Eq:BSSN1},\ref{Eq:BSSN2},\ref{Eq:Lapse},\ref{Eq:BSSN3fo},\ref{Eq:BSSN4fo},\ref{Eq:BSSN5fo},
\ref{Eq:BSSN8},\ref{Eq:BSSN9},\ref{Eq:BSSN10}). It has the form
\begin{equation}
\hat{\partial}_0 u = \alpha {\bf A}^i(u)\partial_i u + F(u),
\label{Eq:QLFO}
\end{equation}
where the matrix-valued functions ${\bf A}^i(u)$, $i=1,2,3$, and the
vector-valued function $F(u)$ depend on $u$ but not their
derivatives. An important point to notice here is that we have not
added any of the constraints variables
(\ref{Eq:BSSNCons1}-\ref{Eq:BSSNCons3}) to the right-hand side (RHS)
of the evolution equations for the extra variables. As a consequence,
the additional constraints, defined by,
\begin{eqnarray}
C^d_k &\equiv&  d_k - 12\partial_k\phi = 0,
\\
C^{\tilde{d}}_{kij} &\equiv& \tilde{d}_{kij} - \partial_k\tilde{\gamma}_{ij} = 0,
\\
C^A_k &\equiv& A_k - \frac{\partial_k\alpha}{\alpha} = 0
\end{eqnarray}
that arise when writing the system as a first order one propagate
independently on whether or not the remaining constraints are
satisfied:
\begin{eqnarray}
\hat{\partial}_0 C^d_k &=& -2\alpha K C^A_k
 + C^d_l\partial_k\beta^l ,\\
\hat{\partial}_0 C^{\tilde{d}}_{kij} &=& -2\alpha\tilde{A}_{ij} C^A_k
 +  C^{\tilde{d}}_{lij}\partial_k\beta^l
 + 2 C^{\tilde{d}}_{kl(i}\partial_{j)}\beta^l - \frac{2}{3}C^{\tilde{d}}_{kij}\partial_l\beta^l ,\\
\hat{\partial}_0 C^A_k &=& -\alpha\frac{\partial F}{\partial\alpha} C^A_k 
 + C^A_l\partial_k\beta^l .
\end{eqnarray}
This means that if initial data is given such that $C^d_k = 0$,
$C^{\tilde{d}}_{kij} = 0$, $C^A_k = 0$ (and suitable boundary
conditions are chosen), these constraints will also be satisfied at
later times and we obtain a solution of the BSSN equations
(\ref{Eq:BSSN1}-\ref{Eq:BSSN5}). This is true even if the initial data
{\em violates} the constraints $H=0$, $M_i=0$, $C_\Gamma^i = 0$ of the
BSSN system.

Having obtained a first order quasi-linear system that yields the same
solutions than the BSSN system (provided that the constraints $C^d_k =
0$, $C^{\tilde{d}}_{kij} = 0$, $C^A_k = 0$ are satisfied initially) we
now analyze for what range of the parameters $m$, $\sigma$ and $\zeta$
the first order system is symmetric hyperbolic. Introducing the
principal symbol ${\bf A}({\bf n}) = {\bf A}^i n_i$ where ${\bf n} =
n_k dx^k$ is any normalized one-form, this means that we have to find
a positive definite matrix ${\bf H} = {\bf H}(u,x^\mu)$ which depends
smoothly on $u$ and the spacetime coordinates $x^\mu$ such that ${\bf
H}{\bf A}({\bf n})$ is symmetric for all $u$, $x^\mu$ and all
normalized one-forms ${\bf n}$\footnote{Notice that
Eq. (\ref{Eq:QLFO}) is equivalent to $\partial_t u = (\alpha {\bf
A}^i(u) + \beta^i)\partial_i u + F(u)$. It is obvious that $(\alpha
{\bf A}^i + \beta^i{\bf I}) n_i$ is symmetrizable if and only if $
{\bf A}^i n_i$ is symmetrizable.}. A necessary condition for this is
that each ${\bf A}({\bf n})$ is diagonalizable and has only real
eigenvalues. So we first analyze the eigenvalue problem
\begin{equation}
\mu\, u = {\bf A}({\bf n})u.
\end{equation}
Explicitly, we have
\begin{eqnarray}
\mu\,\phi &=& 0,
\\
\mu\,\tilde{\gamma}_{ij} &=& 0,
\\
\mu\,\alpha &=& 0,
\\
\mu\, K &=& -A_n\, ,
\\
\mu\, \tilde{A}_{ij} &=& -\frac{1}{2} \tilde{d}_{nij} 
 + \frac{\zeta}{2} \left[ \tilde{d}_{(ij)n} \right]^{TF} 
 + e^{-4\phi}\left[ n_{(i}\tilde{\Gamma}_{j)} - \frac{1}{6} n_{(i}d_{j)} 
  - n_{(i} A_{j)} - \frac{\zeta}{2} n_{(i}\tilde{d}^k_{\; j)k} \right]^{TF}, 
\\
\mu\, \tilde{\Gamma}_i &=& 2(m-1) e^{4\phi} \tilde{A}_{ni} - \frac{4m}{3} n_i K, 
\\
\mu\, d_k &=& -2 n_k K,
\\
\mu\, \tilde{d}_{kij} &=& -2n_k \tilde{A}_{ij}\, ,
\\
\mu\, A_k &=& -2\sigma n_k K,
\end{eqnarray}
where $A_n \equiv A_i n^i$, $\tilde{d}_{nij} = \tilde{d}_{kij} n^k$
etc.  and $\tilde{\Gamma}_i =
\tilde{\gamma}_{ij}\tilde{\Gamma}^j$. Here, and in the following, we
normalize $n_i$ with respect to the three metric $\gamma_{ij}$. A
convenient way for obtaining the nonzero eigenvalues is by deriving a
closed equation for the extrinsic curvature. Introducing $K_{ij} =
e^{4\phi}\tilde{A}_{ij} + \gamma_{ij} K/3$ we obtain
\begin{equation}
\mu^2 K_{ij} = K_{ij} + 2(m-1)n_{(i} K_{j)n} + (1-2m+2\sigma)n_i n_j K 
 + \frac{2}{3}(m-1)\gamma_{ij} (K - K_{nn}).
\end{equation}
In \cite{ST-Live} it was shown that the system is strongly hyperbolic
if the operator on the RHS is diagonalizable and has only {\em
strictly positive} eigenvalues. This is the case if and only if the
squares of the eigenspeeds,
\begin{equation}
\mu_1^2 = 2\sigma, \qquad
\mu_2^2 = \frac{4m-1}{3}\; ,\qquad
\mu_3^2 = m,\qquad
\mu_4^2 = 1,
\label{Eq:SES}
\end{equation}
are strictly positive, that is, if and only if $m > 1/4$ and $\sigma >
0$. Notice that these conditions are independent of $\zeta$ and that
for $\sigma=1/2$ (that is, if $F = \alpha K + F_0(\alpha,x^\mu)$) and
$m=1$ all speeds are one or zero.

In order to find the most general symmetrizer it is convenient to
define $\hat{K}_{ij} = e^{4\phi}\tilde{A}_{ij}$, to decompose
\begin{displaymath}
\tilde{d}_{kij} = -2e^{-4\phi} e_{kij} + \frac{3}{5}\,\tilde{\gamma}_{k(i} b_{j)} - \frac{1}{5}\tilde{\gamma}_{ij} b_k\; ,
\qquad b_j = \tilde{\gamma}^{ki}\tilde{d}_{kij}\; ,
\end{displaymath}
where $e_{kij}$ is completely trace-free, and to replace $\tilde{\Gamma}_i$,
$d_i$, $b_i$ by the combinations
\begin{eqnarray}
v_i &=& \tilde{\Gamma}_i - \frac{1}{6} d_i - A_i - \frac{9\zeta + 6}{20} b_i\; ,
\nonumber\\
w_i &=& \tilde{\Gamma}_i - \frac{1}{6} d_i - A_i + (m-1) b_i\; ,
\nonumber\\
z_i &=& \sigma d_i - A_i\; .
\nonumber
\end{eqnarray}
Here, we assume that $\sigma > 0$ and that $20m + 9\zeta - 14 > 0$
which implies that the transformation is regular. The first condition
is necessary for strong hyperbolicity, and the second one can always
be achieved by choosing the parameter $\zeta$ (which does not appear
in the original BSSN system) to be sufficiently large.

In terms of these variables the non-trivial block of the principal part reads
\begin{eqnarray}
\mu\, K &=& -A_n\, ,
\label{Eq:muK}\\
\mu\, A_i &=& -2\sigma n_i K,
\label{Eq:muAk}\\
\mu\, \hat{K}_{ij} &=& e_{nij} - \zeta e_{(ij)n} + \left[ n_{(i} v_{j)} \right]^{TF},
\label{Eq:muKij}\\
\mu\, e_{kij} &=& n_k\hat{K}_{ij} - \frac{3}{5}\gamma_{k(i} \hat{K}_{j)n} + \frac{1}{5}\gamma_{ij}\hat{K}_{kn}\; ,
\label{Eq:muekij}\\
\mu\, v_i &=& \left( 2m + \frac{9\zeta-14}{10} \right) \hat{K}_{ni} + \left( 2\sigma + \frac{1-4m}{3} \right) n_i K,
\label{Eq:muvi}
\end{eqnarray}
(and $\mu\,\phi = 0$, $\mu\,\tilde{\gamma}_{ij} = 0$, $\mu\,\alpha=0$,
$\mu\, w_i = 0$, $\mu\, z_i = 0$). From this representation of the
principal part it is not difficult to see that the system is symmetric
hyperbolic if and only if
\begin{equation}
4m = 6\sigma + 1, \qquad
\sigma > 0,
\end{equation}
and that in this case a symmetrizer ${\bf H} = {\bf
H}(\gamma^{ij},\sigma,m,\zeta)$ is given by
\begin{eqnarray}
(u^{(1)})^T {\bf H} u^{(2)} &=& 
 \phi^{(1)}\phi^{(2)} + \gamma^{ik}\gamma^{jl}\tilde{\gamma}^{(1)}_{ij}\tilde{\gamma}^{(2)}_{kl} 
 + \alpha^{(1)} \alpha^{(2)} + \gamma^{ij} w^{(1)}_i w^{(2)}_j + \gamma^{ij} z^{(1)}_i z^{(2)}_j
 + 2\sigma K^{(1)} K^{(2)} + \gamma^{ij} A^{(1)}_i A^{(2)}_j 
\nonumber\\
 &+& \gamma^{ik}\gamma^{jl}\hat{K}^{(1)}_{ij}\hat{K}^{(2)}_{kl}
  + \gamma^{kl}\gamma^{ir}\gamma^{js}\left( e^{(1)}_{kij} e^{(2)}_{lrs} - \zeta\, e^{(1)}_{kij} e^{(2)}_{rsl} \right) 
  + \left( 2m + \frac{9\zeta-14}{10} \right)^{-1} \gamma^{ij} v^{(1)}_i v^{(2)}_j\; .
\nonumber 
\end{eqnarray}
In order for ${\bf H}$ to be positive definite we need $-2 < \zeta <
1$. (This can be seen by using the orthogonal decomposition $e_{kij} =
e^s_{kij} + e^a_{kij}$, where $e^s_{kij} = e_{(kij)}$ is totally
symmetric, and by noticing that $e^a_{(ij)k} = -e^a_{kij}/2$.)
Therefore, we have to choose
\begin{equation}
\max\Big\{ -2,1-\frac{10\sigma}{3} \Big\} < \zeta < 1.
\label{Eq:zeta}
\end{equation}
Since $\sigma > 0$ this choice is always possible. Summarizing, we
have shown that our first order system is symmetric hyperbolic if $4m
= 6\sigma + 1 > 1$, $\zeta$ satisfies the inequality (\ref{Eq:zeta})
and if $\sigma$ and $\zeta$ depend smoothly on $u$ and the spacetime
coordinates $x^\mu$. This implies that in those cases the
corresponding initial value problem is well posed. Since the
additional constraints propagate, the same result holds for the BSSN
system with the gauge conditions (a) when $4m = 6\sigma + 1 > 1$ and
$\sigma$ depends smoothly on $u$ and $x^\mu$. Since in this case the
evolution system for the constraint variables can be reduced to
a symmetric hyperbolic system, it follows that the constraints are
satisfied if satisfied initially.  Notice that if $0 < \sigma \leq
1/2$, there are no superluminal speeds. In the next section, we assume
the presence of artificial boundaries and discuss boundary conditions.

\section{Boundary conditions}
\label{Sect:BC}

Consider the BSSN system (\ref{Eq:BSSN1}-\ref{Eq:BSSN5}) on a bounded
domain $\Omega\subset \Real^3$ with smooth boundary $\partial\Omega$.
Consider the slicing condition (\ref{Eq:Lapse}) with $\sigma > 0$ and
choose $m$ such that $4m = 6\sigma + 1$. We also assume that the shift
is a priori specified, and that at the boundary, the shift is
tangential to $\partial\Omega$.

{From} the previous section we know that the BSSN system can be
reduced to a first order symmetric hyperbolic system. For such a
system, the specification of maximally dissipative\footnote{Note that
this terminology has its origin in semigroup theory. Maximality refers
to the fact that generators of semigroups do not have proper
extensions to operators that generate semigroups. Dissipativity of the
operator implies that the spectrum of the generator is contained in
the closed left half-plane. Using analogy from quantum theory,
dissipativity is the condition that all `expectation values' of the
symmetric part of the operator are negative ($\leq 0$). In special
applications those `expectation values' are called `energies' although
in most cases, in particular in General Relativity, they are not
energies in a physical sense, since coordinate, i.e., gauge
dependent. Therefore, calling boundary conditions maximally
dissipative has nothing to do with the boundary conditions dissipating
as much energy to the outside of the computational domain as possible
or the like. Even boundary conditions conserving energies are
maximally dissipative, but, of course, lead to the worst reflections
at the boundaries. The quality of the boundary condition concerning
reflections at the boundary has to be decided by different means, for
instance reflection coefficients for modes incident at the
boundary. Often instead of saying that an operator is dissipative its
negative is referred to as being `accretive'. } boundary conditions
yields a well posed initial-boundary value formulation
\cite{MaxDiss}. Maximally dissipative boundary conditions consist in a
coupling of the ingoing to the outgoing characteristic fields with
respect to the normal ${\bf n}$ to the boundary and some free boundary
data. The characteristic fields with respect to the normal to the
boundary are defined as the projections of $u$ onto the corresponding
eigenspaces of ${\bf A}({\bf n})$.

In order to find the characteristic fields for our first order system,
we define
\begin{eqnarray}
E_{ij} &=& e_{nij} - \zeta\, e_{(ij)n} + \left[ n_{(i} v_{j)} \right]^{TF}
\nonumber\\
 &=& -\frac{1}{2}\, e^{4\phi} \tilde{d}_{nij} 
 + \frac{\zeta}{2}\, e^{4\phi} \left[ \tilde{d}_{(ij)n} \right]^{TF} 
 + \left[ n_{(i}\tilde{\Gamma}_{j)} - \frac{1}{6} n_{(i}d_{j)} 
  - n_{(i} A_{j)} - \frac{\zeta}{2} n_{(i}\tilde{d}^k_{\; j)k} \right]^{TF}, 
\nonumber
\end{eqnarray}
which is trace-free. Equations (\ref{Eq:muKij}, \ref{Eq:muekij},
\ref{Eq:muvi}) imply that
\begin{eqnarray}
\mu\,\hat{K}_{ij} &=& E_{ij}\; ,\\
\mu\, E_{ij} &=& \hat{K}_{ij} + 2(m-1)\left( n_{(i}\hat{K}_{j)n} - \frac{1}{3}\gamma_{ij}\hat{K}_{nn} \right).
\end{eqnarray}
In terms of a triad $e_1$, $e_2$, $e_3$ which is such that $e_1^i =
n^i$, it follows from this and Eqs. (\ref{Eq:muK}, \ref{Eq:muAk}) that
the characteristic fields with respect to the normal $n_i$ that have
nonzero speeds are given by
\begin{eqnarray}
V^{(\pm)} &=& K \mp \mu_1^{-1} A_n\; ,\\
V^{(\pm)}_{nn} &=& \hat{K}_{nn} \pm \mu_2^{-1} E_{nn}\; ,\\
V^{(\pm)}_{nA} &=& \hat{K}_{nA} \pm \mu_3^{-1} E_{nA}\; ,\\
V^{(\pm)}_{AB} &=& \left[ \hat{K}_{AB} \pm E_{AB} \right]^{tf},
\end{eqnarray}
where $A,B$ refer to the triad indices $2$ and $3$, where $[
... ]^{tf}$ denotes the trace-free part with respect to the
two-dimensional metric $\delta_{AB}$ and where $\mu_1$, $\mu_2$,
$\mu_3$ are given by the positive square roots of the expressions in
(\ref{Eq:SES}). A short calculation shows that
\begin{eqnarray}
u^T {\bf H} {\bf A}({\bf n})u
 &=& \sqrt{2\sigma}\,\sigma\left( (V^{(+)})^2 - (V^{(-)})^2 \right)
  + \frac{3\mu_2}{4}\, \left( (V_{nn}^{(+)})^2 - (V_{nn}^{(-)})^2 \right)
\nonumber\\
 &+& \mu_3\,\delta^{AB} \left( V_{nA}^{(+)} V_{nB}^{(+)} - V_{nA}^{(-)} V_{nB}^{(-)} \right)
  + \frac{1}{2}\,\delta^{AC}\delta^{BD}\left( V_{AB}^{(+)} V_{CD}^{(+)} - V_{AB}^{(-)} V_{CD}^{(-)} \right)
\label{Eq:uHAu}
\end{eqnarray}
The maximally dissipative boundary conditions are given as follows:
Let $p\in\partial\Omega$, and let $n_i$ be the unit outward normal
to $\partial\Omega$.  Then, the boundary conditions at $p$ are
\begin{eqnarray}
V^{(+)} &=& a V^{(-)} + G,
\label{Eq:BC1}\\
V^{(+)}_{nn} &=& b V^{(-)}_{nn} + G_{nn}\; ,
\label{Eq:BC2}\\
V^{(+)}_{nA} &=& c_A^B V^{(-)}_{nB} + G_{nA}\; ,
\label{Eq:BC3}\\
V^{(+)}_{AB} &=& d_{AB}^{CD} V^{(-)}_{CD} + G_{AB}\; ,
\label{Eq:BC4}
\end{eqnarray}
where $a$, $b$ are smaller than one in magnitude and the matrices
$c_A^B$ and $d_{AB}^{CD}$ have norm smaller than one, and where $G$,
$G_{nn}$, $G_{nA}$ and $G_{AB}$ are freely specified source functions
(subject to the condition $\delta^{AB} G_{AB} = 0$). In order to
illustrate why these boundary conditions lead to a well posed IBVP,
let us linearize the equations around an arbitrary background. The
resulting equations have the form
\begin{displaymath}
\hat{\partial}_0 v = \alpha {\bf A}^i\partial_i v + {\bf B} v,
\end{displaymath}
where $v$ denotes the perturbation. Defining the energy norm
\begin{displaymath}
{\cal E} = \int_{\Omega} v^T {\bf H} v\, d^3 x,
\end{displaymath}
taking a time derivative, using the symmetries of the matrices ${\bf
H}$ and ${\bf H}{\bf A}^i$ and using Gauss' theorem, we find
\begin{eqnarray}
\frac{d}{dt}\, {\cal E} &=& 2\int_{\Omega} v^T {\bf H}\left[ (\alpha {\bf A}^i + \beta^i)\partial_i v + {\bf B} v \right] d^3 x
\nonumber\\
 &=& \int_{\Omega} \left[ \partial_i\left( v^T \alpha {\bf H}{\bf A}^i v + v^T {\bf H}\beta^i v \right) 
       + v^T\left( {\bf H}{\bf B}
 + {\bf B}^T{\bf H} - \partial_i(\alpha{\bf H}{\bf A}^i + {\bf H}\beta^i) \right)v \right] d^3 x
\nonumber\\
 &\leq& \int_{\partial\Omega} \alpha\, v^T {\bf H}{\bf A}({\bf n}) v\, d^2 x + C E,
\end{eqnarray}
where we have used the fact that the shift is tangential to the
boundary at the boundary and where $C$ is a constant that only depends
on bounds for ${\bf B}$ and ${\bf H}^{-1}\partial_i(\alpha{\bf H}{\bf
A}^i + {\bf H}\beta^i)$. If the boundary conditions are homogeneous,
i.e. if $G = 0$, $G_{nn} = 0$, $G_{nA} = 0$, $G_{AB} = 0$,
Eqs. (\ref{Eq:uHAu},\ref{Eq:BC1},\ref{Eq:BC2},\ref{Eq:BC3},\ref{Eq:BC4})
immediately imply that the boundary integral is negative or zero, and
we obtain the energy estimate ${\cal E}(t) \leq \exp(Ct) {\cal
E}(0)$. If the boundary conditions are inhomogeneous one can bound
${\cal E}(t)$ by ${\cal E}(0)$ and the $L2$-norm of the boundary data
\cite{KL, GKO-Book}. These energy estimates play a key role in proofs
for well posedness. These proofs can be generalized to quasi-linear
symmetric hyperbolic systems, see for instance \cite{MaxDiss}.

Therefore, the boundary conditions
(\ref{Eq:BC1},\ref{Eq:BC2},\ref{Eq:BC3},\ref{Eq:BC4}) lead to a well
posed initial-boundary value formulation. Since the shift is
tangential to the boundary at $\partial\Omega$, the additional
constraints propagate as before, and thus the same boundary conditions
applied to the equations (\ref{Eq:BSSN1}-\ref{Eq:BSSN5}), where we
perform the replacements (\ref{Eq:ExtraVars}), yields a well posed
initial-boundary formulation for the BSSN system. In particular,
choosing $a=b=0$, $c_A^B=0$, $d_{AB}^{CD} = 0$, and setting the source
functions $G$, $G_{nn}$, $G_{nA}$, $G_{AB}$ to zero, corresponds to
Sommerfeld-type boundary conditions, in the sense that these
conditions are algebraic conditions for the first order systems which
are perfectly absorbing for plane waves of normal incidence to the
boundary in the frozen coefficient approximation. Explicitly, we
obtain the six boundary conditions
\begin{eqnarray}
K - \frac{1}{\sqrt{2\sigma}\, \alpha}\, n^i\partial_i\alpha &=& 0, 
\label{Eq:BCSF1}\\
n^i n^j\left[ e^{4\phi}\tilde{A}_{ij} + \frac{\sqrt{3}}{\sqrt{4m-1}}\, E_{ij} \right] &=& 0, 
\label{Eq:BCSF2}\\
n^i e^j_A \left[ e^{4\phi}\tilde{A}_{ij} + \frac{1}{\sqrt{m}}\, E_{ij} \right] &=& 0, 
\qquad A=2,3,
\label{Eq:BCSF3}\\
\left[ e^i_A e^j_B - \frac{1}{2}\, \delta_{AB}\delta^{CD} e^i_C e^j_D \right]
\left[ e^{4\phi}\tilde{A}_{ij} + E_{ij} \right] &=& 0,
\qquad A,B=2,3,
\label{Eq:BCSF4}
\end{eqnarray}
where
\begin{displaymath}
E_{ij} = -\frac{1}{2}\, e^{4\phi} n^k\partial_k \tilde{\gamma}_{ij} 
 + \left[ n_{(i}\tilde{\Gamma}_{j)} - 2 n_{(i} \partial_{j)}\phi 
 - \frac{1}{\alpha}\, n_{(i} \partial_{j)}\alpha 
 + \frac{\zeta}{2} \left(  e^{4\phi} n^k\partial_{(i}\tilde{\gamma}_{j)k}
 - n_{(i}\tilde{\gamma}^{rs}\partial_{|r|} \tilde{\gamma}_{j)s} \right) \right]^{TF}, 
\end{displaymath}
where $n^i$ is the unit outward normal to the boundary and the vectors
$e_2$ and $e_3$ must be chosen such that $n^i$, $e^i_2$, $e^i_3$ form
a triad with respect to the three metric $\gamma_{ij}$, and $n_i =
\gamma_{ij} n^j$. The vectors $e_2$ and $e_3$ are unique up to a
rotation; such a rotation does not alter the boundary conditions. The
parameter $\zeta$ has to be chosen such that the inequality
(\ref{Eq:zeta}) is satisfied. The boundary conditions
(\ref{Eq:BCSF1},\ref{Eq:BCSF2},\ref{Eq:BCSF3},\ref{Eq:BCSF4}) can be
generalized to inhomogeneous conditions by replacing the zeroes on
their right-hand sides by freely specifiable source functions $G$,
$G_{nn}$, $G_{nA}$, $G_{AB}$. If the solution is known in a
neighborhood of the boundary, one can compute these source functions
by evaluating the left-hand sides of
Eqs. (\ref{Eq:BCSF1},\ref{Eq:BCSF2},\ref{Eq:BCSF3},\ref{Eq:BCSF4}).
Notice that the occurrence of the parameter $\zeta$, which does not
appear in the BSSN system, in the boundary conditions has its origin
in the $\zeta$-dependence of the unphysical energy ${\cal E}$ defined
by the symmetrizer.

\section{Strong hyperbolicity with a dynamical shift}
\label{Sect:LiveShift}

Here we consider the BSSN equations (\ref{Eq:BSSN1}-\ref{Eq:BSSN5})
with the live gauge conditions (b), see section \ref{Sect:BSSNEq}. In
this case one could proceed as in the frozen shift case and introduce
the shift and its first derivatives (with respect to time and space)
as extra variables. One obtains a first order system that is
equivalent to the original system provided that the additional
constraints are satisfied. Unfortunately, we did not succeed in
finding a symmetrizer for the resulting first order system. Our goal
in this section, therefore, is more modest: We show that the BSSN
system with the live gauge conditions is strongly hyperbolic and so
prove that the resulting Cauchy problem (in the absence of boundaries)
is well posed. A related analysis for a different form of the system
has been performed in \cite{BonaPal}.

For differential equations that are not first order, a definition of
strong hyperbolicity has recently be given in \cite{KO, NOR} that does
not require the introduction of extra variables (nor extra
constraints). It is based on pseudo-differential calculus. The
intuitive idea behind this definition is to freeze the coefficients in
the differential equations at some fixed point and to analyze the
resulting linear constant coefficient problem by means of a Fourier
transformation in space. In our case, the frozen coefficient problem
is given by
\begin{eqnarray}
\hat{\partial}_0\hat{\phi} &=& -\frac{\alpha}{6}\,\hat{K} + \frac{i}{6}\omega_k\hat{\beta}^k, 
\nonumber\\
\hat{\partial}_0\hat{\gamma}_{rs} &=& -2\alpha\hat{A}_{rs} 
 + 2i\tilde{\gamma}_{k(r}\omega_{s)}\hat{\beta}^k - \frac{2i}{3}\tilde{\gamma}_{rs}\omega_k\hat{\beta}^k ,
\nonumber\\
\hat{\partial}_0\hat{K} &=& e^{-4\phi}\tilde{\gamma}^{kl}\omega_k\omega_l\hat{\alpha} + l.o.,
\nonumber\\
\hat{\partial}_0 \hat{A}_{rs} &=& \alpha e^{-4\phi} 
  \left[ \frac{1}{2}\tilde{\gamma}^{kl}\omega_k\omega_l\hat{\gamma}_{rs} + i\tilde{\gamma}_{k(r}\omega_{s)}\hat{\Gamma}^k
  + 2\omega_r\omega_s\hat{\phi} + \omega_r\omega_s\frac{\hat{\alpha}}{\alpha} \right]^{TF}
 + l.o.,
\nonumber\\
\hat{\partial}_0\hat{\Gamma}^s &=& -\tilde{\gamma}^{kl}\omega_k\omega_l\hat{\beta}^s
 - \frac{1}{3} \tilde{\gamma}^{rs}\omega_r\omega_k\hat{\beta}^k 
 + 2\alpha\left[ i(m-1)\omega_k\hat{A}^{ks} - \frac{2i m}{3}\tilde{\gamma}^{rs}\omega_r K \right]
 + l.o.,
\nonumber\\
\hat{\partial}_0\hat{\alpha} &=& -\alpha^2 f(\alpha,\phi,x^\mu)\hat{K} + l.o.,
\nonumber\\
\hat{\partial}_0\hat{\beta}^s &=& \alpha^2 G(\alpha,\phi,x^\mu) \hat{B}^s,
\nonumber\\
\hat{\partial}_0\hat{B}^s &=&  e^{-4\phi} H(\alpha,\phi,x^\mu)\left\{ -\tilde{\gamma}^{kl}\omega_k\omega_l\hat{\beta}^s
 - \frac{1}{3} \tilde{\gamma}^{rs}\omega_r\omega_k\hat{\beta}^k 
 + 2\alpha\left[ i(m'-1)\omega_k\hat{A}^{ks} - \frac{2i m'}{3}\tilde{\gamma}^{rs}\omega_r K \right] \right\}
 + l.o.,
\nonumber
\end{eqnarray}
where a hat denotes the Fourier transformation in space,
$\hat{\phi}(\omega) = \int \phi(x)\exp(-i\omega\cdot x)\, d^3 x$, and
$l.o.$ denotes terms that depend on lower order spatial
derivatives. Here, we have also allowed for a parameter $m'$ that is
different than $m$ in the evolution equation for $B^i$. We can rewrite
this as a first order system in $t$ and $\omega_i$ by writing
$\omega_i = |\omega| n_i$, $|\omega| =
\sqrt{\gamma^{kl}\omega_k\omega_l}$, and introducing the variables
\begin{eqnarray}
\hat{\varphi} &=& i|\omega| \hat{\phi},
\nonumber\\
\hat{h}_{rs} &=& \frac{i|\omega|}{2} e^{4\phi}\hat{\gamma}_{rs}\; ,
\nonumber\\
\hat{a} &=& i\alpha^{-1} |\omega|\hat{\alpha},
\nonumber\\
\hat{b}_s &=& i\alpha^{-1} |\omega|\gamma_{rs}\hat{\beta}^r ,
\nonumber\\
\hat{k}_{rs} &=& e^{4\phi}\hat{A}_{rs}\; ,
\nonumber\\
\hat{\Gamma}_s &=& \tilde{\gamma}_{rs}\hat{\Gamma}^r,
\nonumber\\
\hat{B}_s &=& \gamma_{rs}\hat{B}^r.
\nonumber
\end{eqnarray}
In terms of the variables we obtain a first order pseudo-differential system of
the form
\begin{displaymath}
\partial_t\hat{u} = i|\omega|\left( \alpha {\bf P}({\bf n}) + \beta^i n_i \right)\hat{u} + l.o.,
\end{displaymath}
where $\hat{u} =
(\hat{\varphi},\hat{h}_{ij},\hat{K},\hat{k}_{ij},\hat{a},\hat{b}_i,\hat{\Gamma}_i,\hat{B}_i)^T$.
The system is strongly hyperbolic if there exists a positive definite
Hermitian matrix ${\bf H}(x^\mu,u,{\bf n})$ which is smooth in all its
entries such that ${\bf H} {\bf P}$ is symmetric. A necessary
condition for this is that ${\bf P}$ is diagonalizable and has only
real eigenvalues. Therefore, we first consider the eigenvalue problem
$\mu\,\hat{u} = {\bf P}({\bf n})\hat{u}$; explicitly
\begin{eqnarray}
\mu\,\hat{\varphi} &=& -\frac{1}{6}\,\hat{K} + \frac{1}{6}\,\hat{b}_n\; ,
\nonumber\\
\mu\,\hat{h}_{rs} &=& -\hat{k}_{rs} + \left[ n_{(r}\hat{b}_{s)} \right]^{TF},
\nonumber\\
\mu\,\hat{K} &=& -\hat{a},
\nonumber\\
\mu\,\hat{k}_{rs} &=& -\hat{h}_{rs} + \left[ n_{(r}\hat{\Gamma}_{s)} - 2n_r n_s\hat{\varphi} - n_r n_s\hat{a} \right]^{TF},
\nonumber\\
\mu\,\hat{a} &=& -f \hat{K},
\nonumber\\
\mu\,\hat{b}_s &=& G \hat{B}_s,
\nonumber\\
\mu\,\hat{\Gamma}_s &=& \hat{b}_s + \frac{1}{3}n_s \hat{b}_n + 2(m-1)\hat{k}_{ns} - \frac{4m}{3} n_s\hat{K},
\nonumber\\
\mu\,\hat{B}_s &=& H\left[ \hat{b}_s + \frac{1}{3}n_s \hat{b}_n + 2(m'-1)\hat{k}_{ns} - \frac{4m'}{3} n_s\hat{K} \right],
\nonumber
\end{eqnarray}
where $\hat{b}_n = \gamma^{rs} n_r\hat{b}_s$ and $\hat{k}_{nj} =
\gamma^{rs} n_r\hat{k}_{sj}$. A careful analysis reveals that the
matrix on the RHS has the eigenvalues $0$, $\pm\mu_1$, $\pm\mu_2$,
$\pm\mu_3$, $\pm\mu_4$, $\pm\mu_5$ where
\begin{displaymath}
\mu_1 = \sqrt{f}, \qquad
\mu_2 = \sqrt{\frac{4m-1}{3}}\; ,\qquad
\mu_3 = \sqrt{m},\qquad
\mu_4 = 1, \qquad
\mu_5 = \sqrt{GH}\; , \qquad
\mu_6 = \sqrt{\frac{4GH}{3}}\; .
\end{displaymath}
Therefore, we need $m > 1/4$, $f > 0$ and $GH > 0$. (If $G=H=0$ the
equation for the shift decouples, and we are back in the case
considered in the previous section). Furthermore, it turns out that
the matrix is diagonalizable only if $4GH \neq 3f$ and provided that
$m'=1$ if $m=GH$ or $4GH = 4m-1$. In the remaining cases the system is
only weakly hyperbolic which, in the nonlinear case, can lead to
exponential growth with arbitrarily small growth time. Introducing
the factors
\begin{eqnarray}
\Omega_1 &=& \frac{4GH}{3f - 4GH}\; ,
\nonumber\\
\Omega_2 &=& \frac{6(m'-1)}{4m-1-4GH}\; ,\quad \hbox{if $4m-1 \neq 4GH$ and $\Omega_2$ arbitrary otherwise},
\nonumber\\
\Omega_3 &=& \frac{2(m'-1)GH}{m-GH}\; , \quad \hbox{if $m \neq GH$ and $\Omega_3$ arbitrary otherwise},
\nonumber
\end{eqnarray}
the eigenfields can be expressed as
\begin{eqnarray}
Z_0 &=& 8m\hat{\varphi} - 2(m-1)\hat{h}_{nn} - \hat{\Gamma}_n\; ,
\nonumber\\
Z_i &=& H \left[ 2(m-m')\hat{h}_{ni} + m'\hat{\Gamma}_i \right] - m\hat{B}_i\; ,
\nonumber\\
V^{(\pm)} &=& \hat{K} \mp \mu_1^{-1}\hat{a}\; ,
\nonumber\\
V^{(\pm)}_{nn} &=& \hat{k}_{nn} - \frac{2\hat{K}}{3} \mp \mu_2^{-1}
  \left( \hat{h}_{nn} - \frac{2}{3}\hat{\Gamma}_n + \frac{4}{3}\hat{\varphi} \right),
\nonumber\\
V^{(\pm)}_{nA} &=& \hat{k}_{nA} \mp \mu_3^{-1}\left( \hat{h}_{nA} - \frac{1}{2}\hat{\Gamma}_A \right),
\nonumber\\
V^{(\pm)}_{AB} &=& \left[ \hat{k}_{AB} \mp \mu_4^{-1}\hat{h}_{AB} \right]^{tf},
\nonumber\\
V^{(\pm)}_A &=& \hat{b}_A - \Omega_3\hat{k}_{nA} 
  \pm \mu_5^{-1} \left[ G\hat{B}_A + \Omega_3\left( \hat{h}_{nA}-\frac{1}{2}\hat{\Gamma}_A \right) \right],
\nonumber\\
V^{(\pm)}_n &=& \hat{b}_n + \Omega_1\hat{K} - \Omega_2\left( \hat{k}_{nn} - \frac{2\hat{K}}{3} \right) 
  \pm \mu_6^{-1} \left[ G\hat{B}_n - \Omega_1\hat{a} + \Omega_2\left(\hat{h}_{nn}-\frac{2}{3}\hat{\Gamma}_n + \frac{4}{3}\hat{\varphi} \right) \right],
\nonumber
\end{eqnarray}
where the components $n$, $A=2,3$, refer to triad indices as
before. The matrix ${\bf H}$ which symmetrizes ${\bf P}$ can now be
built by summing over the square of the eigenfields. It is smooth in
the one-form $n_i$. In order to see this one uses, for example,
$\hat{h}_{nA}\hat{h}^{nA} = n^i n^k(\gamma^{jl} - n^j
n^l)\hat{h}_{ij}\hat{h}_{kl}$. It is also smooth in the other
variables provided that $\Omega_1$, $\Omega_2$ and $\Omega_3$ can be
chosen smoothly and are bounded. A simple possibility of achieving
this is by choosing $m=m'=1$ and $f = \kappa G H$ with $\kappa$ a
constant that is unequal $4/3$. The pseudo-differential calculus shows
that in these cases the full nonlinear Cauchy problem is well posed.
Since the evolution system for the constraint variables can be reduced
to a symmetric hyperbolic system if $m > 1/4$ it follows that the
constraints are satisfied if satisfied initially.

\section{Conclusion}
\label{Sect:Conclusion}

We discussed some mathematical aspects of the BSSN system which is
currently used by several groups in numerical relativity. In
particular, we derived a well posed initial-boundary value formulation
of the BSSN system with a Bona-Mass{\'o} like slicing condition for
the lapse and a frozen shift. This is achieved by introducing extra
variables and recasting the evolution equations into a first order
symmetric hyperbolic system, for which maximally dissipative boundary
conditions are specified. The introduction of extra fields brings
additional constraints, and the original BSSN system and the first
order symmetric hyperbolic system derived in this article are only
equivalent if these constraints are satisfied. However, we showed that
the associated constraint variables obey a closed evolution system
that is {\em independent} of the other constraints. Moreover, by
choosing the shift to be tangential to the boundary, these additional
constraints propagate tangentially to the boundary. This implies that
they are satisfied everywhere at later times if satisfied initially,
even if the other constraints are violated. This allows us to return
to the second order system and to conclude that the BSSN system with
the specified boundary conditions is well posed; in particular, unique
solutions local in time exist, and depend continuously on the initial
and boundary data. To our knowledge, the specified (six) boundary
conditions
(\ref{Eq:BCSF1},\ref{Eq:BCSF2},\ref{Eq:BCSF3},\ref{Eq:BCSF4}) have not
yet appeared in the literature.
\newline
\linebreak In general, the boundary conditions derived in this article
are not compatible with the constraints of the BSSN system. They can
feed in some constraint violating modes. Nevertheless, they are
consistent with the evolution equations and constitute a first step
towards improving numerical evolutions of the BSSN system. In
particular, the present analysis offers the possibility to construct
constraint-preserving boundary conditions \cite{CPBC} in the
linearized case, following the lines of \cite{CPSTR, CS,
GMG2}. Furthermore, the derivation of the symmetrizer and the energy
estimate presented in section \ref{Sect:BC} should be useful as a
guidance principle to construct discretizations schemes that guarantee
numerical stability at least at the linearized level \cite{GKO-Book,
Olsson, RK, LSU-SumPart}.
\newline
\linebreak We have also considered dynamical gauge conditions for
lapse and shift and obtained a class of second order evolution
equations which can be shown to be strongly hyperbolic using
pseudo-differential calculus. For these systems, one can show well
posedness of the initial value problem. Here, the presence of
boundaries has not been considered. To derive boundary conditions in
the case of a finite domain one could proceed as follows: First,
derive a first order system by introducing extra variables as
described at the beginning of section \ref{Sect:LiveShift}. Next,
consider the matrix ${\bf A}({\bf n})$ multiplying the derivatives
normal to the boundaries. In case this matrix is diagonalizable, to
every strictly positive eigenvalue of ${\bf A}({\bf n})$ there
corresponds a Sommerfeld-type outgoing boundary condition given by the
condition of a vanishing projection of the field vector onto the
corresponding eigenspace. This corresponds to setting zero the
incoming characteristic fields with respect to the direction which is
normal to the boundary. Finally, well posedness of the
initial-boundary value problem in a suitable Hilbert space has to be
proved\footnote{Notice that setting to zero the incoming
characteristic fields for systems that are strongly but not symmetric
hyperbolic does not always lead to a well posed problem, see \cite{CS}
for a counterexample.}. Necessary conditions for well posedness can be
obtained by using the method of Laplace transformation, see, for
example, \cite{KL, GKO-Book}. The derivation of boundary conditions in
the dynamical shift case is beyond the scope of the present work.
\newline
\linebreak The gauge conditions considered here differ from the ones
used in \cite{AEI} for numerical simulations only by the replacement
$\partial_t\mapsto \partial_t - \beta^j\partial_j$, which leads to a
simpler principal part and makes it more amendable to analyze the
algebraic conditions that guarantee symmetric or strong hyperbolicity.
Preliminary investigations of the problem without this replacement
have been done in \cite{BC-preprint} where one of the Sommerfeld-type
conditions has already been computed. The structure of this condition
is more complicated than the conditions derived in this article.

%%%%%%%%%%%%%%%%%%%%%%%%%%%%%%%%%%%%%%%%%%%%%%%%%%%%%%%%%%%%%%
\section{Acknowledgements}
%%%%%%%%%%%%%%%%%%%%%%%%%%%%%%%%%%%%%%%%%%%%%%%%%%%%%%%%%%%%%%

We wish to thank M. Alcubierre, G. Allen, G. Calabrese, C. Gundlach,
L. Lehner, G. Nagy, E. Seidel, and M. Tiglio for useful comments and
discussions.  This work was supported by the Center for Computation \&
Technology at Louisiana State University, by the Max-Planck-Institut
for Gravitational Physics, by grant NSF-PHY0244335, NASA-NAG5-13430
and by funds from the Horace Hearne Jr. Laboratory for Theoretical
Physics.

%%%%%%%%%%%%%%%%%%%%%%%%%%%%%%%%%%%%%%%%%%%%%%%%

\end{document}